\documentclass{dcds-B}
\usepackage{amsmath}

\def\phi{\varphi} \def\epsilon{\varepsilon} \def\u{\boldsymbol}

\title[Renormalization for cubic invariant tori]
      {Renormalization for cubic frequency invariant tori in
    Hamiltonian systems with two degrees of freedom}
\author[C. Chandre]{}
\email{chandre@cns.physics.gatech.edu}

\begin{document}

\maketitle

\centerline{\scshape   C. Chandre}
 \medskip

  {\footnotesize \centerline{ Center for Nonlinear Science }
  \centerline{ School of Physics, Georgia Institute of Technology }
\centerline{ Atlanta, GA 30332-0430,    U.S.A. } }
 \medskip


 \medskip

\begin{abstract}
  We consider the break-up of invariant tori in Hamiltonian systems
  with two degrees of freedom with a frequency which belongs to a cubic
  field. We define and construct renormalization-group transformations
  in order to determine the threshold of the break-up of these tori. A
  first transformation is defined from the continued fraction
  expansion of the frequency, and a second one is defined with a
  fixed frequency vector in a space of Hamiltonians with three
  degrees of freedom.
\end{abstract}

\section{Introduction}
\label{sec:intro}

\indent We consider the dynamics of a particle in a potential $V$
described by the following Hamiltonian~:
\begin{equation}
\label{eqn:1.5}
H(p,x,t)=\frac{1}{2}p^2 +V(x,t),
\end{equation}
where $V$ is $2\pi$-periodic in $x$ and $t$.
For $V=0$, the phase space is foliated with invariant tori.
For $V$ sufficiently
small and regular, the KAM theorem states that most of these invariant
tori persist.\\
\indent  The
purpose of this paper is to construct renormalization transformations in order to study the
persistence of
invariant
tori with a given cubic frequency $\omega$ smaller
than one, satisfying
$$
\omega^3=a \omega^2+b\omega +1,
$$
where $(a,b)\in {\Bbb Z}^2$.\\ 

\indent For Hamiltonian systems with two and three 
degrees of freedom, renormalization-group transformations in the framework of 
Ref.~\cite{koch99} have been successfully applied for the determination of the 
threshold of the break-up of invariant tori~\cite{chan98b,chan00d,chan01a}.
They have also been used for the analysis of the properties
of critical invariant tori (at the threshold of the
break-up)~\cite{chan98b,abad98,chan00a,koch99b,koch01}. \\
\indent For two degrees of freedom and for reduced quadratic irrational 
frequencies, renormalization transformations
have been defined with a fixed frequency vector, i.e.\ at each
iteration of the transformation, the frequency vector of the torus was
unchanged. This feature was based on the fact that the continued
fraction expansion of the frequency was periodic. For cubic
(and more generally for non-quadratic irrational) frequencies, this is
not the case, and thus ergodic renormalization
was constructed from the entries of the continued fraction of the
frequency. This transformation allows the accurate determination of critical
couplings by comparison with Greene's 
criterion~\cite{gree79} and Laskar's frequency map 
analysis~\cite{lask99} (see Ref.~\cite{chan01R}). \\ 
\indent In order to define a renormalization with a fixed frequency vector, one 
has to add one degree of freedom to the system as explained below. For 
degenerate Hamiltonians with three degrees of freedom, a renormalization 
transformation has been defined for the spiral mean frequency vector. This 
transformation allows one to determine accurately the thresholds of break-up of 
tori by comparison with Laskar's Frequency Map Analysis~\cite{chan00d}.\\ 
\indent In this article, we 
propose to define and study numerically two renormalization transformations for 
invariant tori with cubic frequency~: one based on the continued fraction of the 
frequency acting within a space of Hamiltonians with two degrees of freedom, and 
the other one with a fixed frequency vector acting within a space of degenerate 
Hamiltonians with three degrees of freedom.\\ 
\indent We apply both renormalization 
methods to the numerical determination of the threshold of invariant tori of a 
given one-parameter family of Hamiltonians, Escande's paradigm 
model~\cite{esca85} which is a forced pendulum model. We consider three cases~: 
the spiral mean frequency, the Tribonacci mean frequency, and the
$\tau$-mean frequency.

\section{Renormalization from the continued fraction expansion}
\label{sec:2d}

\indent In order to investigate the torus with frequency $\omega$, we map the
time-dependent
Hamiltonian~(\ref{eqn:1.5}) into a time-independent Hamiltonian with
two degrees of freedom, written in terms of actions
$\u{A}=(A_1,A_2)\in {\Bbb R}^2$ and angles
$\u{\varphi}=(\varphi_1,\varphi_2)\in{\Bbb T}^2$~:
$$
H(A_1,A_2,\varphi_1,\varphi_2)=\frac{1}{2}A_1^2-A_2+V(\varphi_1,-\varphi_2).
$$
\indent The two-dimensional invariant torus is now located around $A_1=
\omega$. We shift the action $A_1$ in order to locate this torus
around $A_1=0$. The Hamiltonian becomes~:
$$
H(\u{A},\u{\varphi})=\u{\omega}\cdot\u{A}+\frac{1}{2}A_1^2
+V(\varphi_1,-\varphi_2),
$$
where the frequency vector of the considered two dimensional torus is now
$\u{\omega}=(\omega,-1)$.\\

\indent The renormalization we define relies upon the continued fraction 
expansion of $\omega$~:
$$ \omega=\frac{1}{a_0+\frac{1}{a_1+\cdots}} \equiv [a_0,a_1,\ldots].$$

\indent This transformation will act within a space of
 Hamiltonians $H$ of the form
\begin{equation}
\label{eqn:HRG}
H(\u{A},\u{\varphi})=\u{\omega}\cdot\u{A}+V(\u{\Omega}\cdot\u{A},\u{\varphi}),
\end{equation}
where $\u{\Omega}=(1,\alpha)$ is a vector not parallel to the
frequency vector $\u{\omega}$. We assume that Hamiltonian~(\ref{eqn:HRG})
satisfies a non-degeneracy condition~: If we expand $V$ into
$$
V(\u{\Omega}\cdot\u{A},\u{\varphi})=\sum_{\scriptstyle \u{\nu} \in \mathbb{Z}^2 \atop
k\geq 0} V^{(k)}_{\u{\nu}}
(\u{\Omega}\cdot\u{A})^k e^{i\u{\nu}\cdot\u{\varphi}},
$$
we assume that $V^{(2)}_{\u{0}}$ is non-zero.\\
\indent The transformation contains essentially two parts~: a rescaling and an 
elimination procedure. We follow the renormalization scheme defined in 
Ref.~\cite{chan00a}. Similar renormalization transformations have been defined 
in Refs.~\cite{lope01a,lope01b} for vector fields.   \\

{\bf (1)} \textit{Rescaling}~: The first part of the transformation is
 composed by a shift of the resonances, a rescaling of time and a
 rescaling of the actions. It acts on a Hamiltonian $H$ as $H'=H\circ
 {\mathcal T}$ (see Ref.~\cite{chan00a} for details)~:
\begin{equation}
H'(\u{A},\u{\varphi})=\lambda\omega^{-1} H\left(-\frac{1}{\lambda}N_{a}\u{A},
-N_{a}^{-1}\u{\varphi}\right),
\end{equation}
with
\begin{equation}
\label{eqn:rescaling}
\lambda=2\omega^{-1} (a+\alpha)^2 V^{(2)}_{\u{0}},
\end{equation}
and
\begin{equation}
  \label{eq:matrix2d}
  N_{a}=\left(\begin{array}{cc} a & 1\\ 1 & 0 \end{array}\right),
\end{equation}
and $a$ is the integer part of $\omega^{-1}$ (the first entry in the continued fraction expansion).
This change of coordinates is a generalized (far from identity) canonical transformation
and the rescaling $\lambda$ is chosen in order to ensure a normalization condition (the
quadratic term in the actions has a mean value equal to 1/2).
For $H$ given by Eq.\ (\ref{eqn:HRG}), this expression becomes
\begin{equation}
H'(\u{A},\u{\varphi})=\u{\omega}'\cdot\u{A}+
\sum_{\u{\nu},k} V^{'(k)}_{\u{\nu}}
(\u{\Omega}'\cdot\u{A})^k e^{i\u{\nu}\cdot\u{\varphi}},
\end{equation}
where
\begin{eqnarray}
  \label{eq:renexp}
  && \u{\omega}'=(\omega',-1) \; \mbox{ with }
  \omega'=\omega^{-1}-a,\\
  && \u{\Omega}'=(1,\alpha) \; \mbox{ with } \alpha'=1/(a+\alpha),\\
  && V^{'(k)}_{\u{\nu}}=r_k V_{-N\u{\nu}}^{(k)} \; \mbox{ with }
  r_k=(-1)^k 2^{1-k}\omega^{k-2}(a+\alpha)^{2-k} \left(
  V^{(2)}_{\u{0}}\right) ^{1-k}.
\end{eqnarray}
\indent We notice that
the frequency of the torus is changed according to the Gauss map~:
\begin{equation}
\label{eqn:gauss}
\omega \mapsto \omega'=\omega^{-1}-\left[ \omega^{-1}\right],
\end{equation}
where $\left[\omega^{-1} \right]$ denotes the integer part of $\omega^{-1}$.
Expressed in terms of
the continued fraction expansion, it corresponds to a shift to the
left of the entries
$$
\omega=[a_0,a_1,a_2,\ldots]\mapsto \omega'=[a_1,a_2,a_3,\ldots].
$$

\textit{Remark:} After renormalization (one or more iterations) the frequency 
vector $\u{\omega}=(\omega, -1)$ is changed  into $\u{\omega}'=\mu M\u{\omega}$ 
with $M\in GL_2(\mathbb{Z})$. In order to define a renormalization with a 
fixed frequency vector $\u{\omega}'=\u{\omega}$, the frequency $\omega$ has to 
be a quadratic irrational. For a cubic frequency (e.g., satisfying 
$\omega^3=a\omega^2+b\omega+1$), such renormalization cannot be defined with a 
fixed frequency vector. The idea is to add one more dimension and consider 
frequency vectors of the form $(\omega, -1,\omega_1)$. If $\omega_1$ is 
carefully chosen, there exists $\mu$ and $M\in GL_3(\mathbb{Z}) $ such that 
$\u{\omega}=\mu M\u{\omega}$, and then one can define a similar renormalization 
with the exception that now there is an additional degree of freedom 
corresponding to the additional frequency $\omega_1$.\\

{\bf (2)} \textit{Elimination}~: The second step is a (connected to identity) canonical
transformation ${\mathcal U}$ that eliminates the non-resonant modes of
the perturbation in $H'$.\\
Following Ref.~\cite{chan00a}, we consider the
set $I^- \subset \mathbb{Z}^2$ of non-resonant modes as the set of integer
vectors $\u{\nu}=(\nu_1,\nu_2)$ such that $|\nu_2|> |\nu_1|$.  Other choices of 
resonant modes can be done without affecting the convergence and speed of the 
algorithm (see Ref.~\cite{chan01R} for a detailed discussion). A mode which is 
not an element of $I^-$, will be called resonant. The canonical transformation 
${\mathcal U}$ is such that $H''=H'\circ {\mathcal U}$ does not have any 
non-resonant mode, i.e.\ it is defined by the following equation~: 
\begin{equation}   \label{eq:proj}
  {\mathbb{I}}^-(H'\circ {\mathcal U})=0,
\end{equation}
where ${\mathbb{I}}^-$ is the projection operator on the non-resonant
modes; it acts on a Hamiltonian~(\ref{eqn:HRG}) as~:
$$
{\mathbb{I}}^- H=\sum_{\scriptstyle \u{\nu}\in I^- \atop k\geq 0} V_{\u{\nu}}^{(k)}
(\u{\Omega}\cdot\u{A})^k e^{i\u{\nu}\cdot\u{\varphi}}.
$$
\indent We solve Eq.~(\ref{eq:proj}) by a Newton's method following
Ref.~\cite{chan98b,chan00a}. We notice that in
Refs.~\cite{chan98b,chan00a} a renormalization procedure acting within
a space of quadratic Hamiltonians in the actions has been constructed.
 This renormalization is the one used in this paper for numerical
purposes.
This transformation gave the same accuracy than the renormalization defined for Hamiltonians in power
series in the actions for
the numerical
computation of the parameters that characterize the break-up of invariant
tori.\\

These rescaling and elimination procedures define the renormalization transformation acting
on a space of Hamiltonians~(\ref{eqn:HRG}) with two degrees of freedom
as $H''={\mathcal R}(H)=H\circ {\mathcal T} \circ{\mathcal U}$. The main
point is that the frequency $\omega$ is changed according to the
Gauss map~(\ref{eqn:gauss}) each time we iterate the transformation.

\section{Renormalization with a fixed frequency vector}
\label{sec:3d}

\indent In order to study a torus with frequency $\omega$, we consider another
frequency $\omega_1$ such that $(1,\omega,\omega_1)$ is a complete
integral basis of the
cubic field to which $\omega$ belongs to. We consider the invariant torus with frequency vector
$\u{\omega}=(\omega,-1,\omega_1)$. The Hamiltonian~(\ref{eqn:1.5}) is
mapped into the time independent Hamiltonian with three degrees of
freedom written in terms of actions
$\u{A}=(A_1,A_2,A_3)\in {\Bbb R}^3$ and angles
$\u{\varphi}=(\varphi_1,\varphi_2,\varphi_3)\in{\Bbb T}^3$~:
$$
H(A_1,A_2,A_3,\varphi_1,\varphi_2,\varphi_3)=\frac{1}{2}A_1^2-A_2+V(\varphi_1,-\varphi_2).
$$
\indent We shift the action $A_1$ by $\omega$ and we add $\omega_1 A_3$ since
$A_3(t)$ is constant ($H$ does not depend on $\varphi_3$). The Hamiltonian becomes~:
$$
H(\u{A},\u{\varphi})=\u{\omega}\cdot\u{A}+\frac{1}{2}A_1^2 + V(\varphi_1,-\varphi_2).
$$
\indent We assume that $\u{\omega}$ is an
eigenvector of a matrix $N$ with integer coefficients,
determinant $\pm 1$, and a real eigenvalue $\mu$ of modulus smaller
than one. For instance, if $\omega_1=\omega^2$,
$$
N=\left( \begin{array}{ccc} 0 & 0 &1 \\ -1 & 0 & 0 \\ b & -1 & a
  \end{array} \right),
$$
and $\mu=\omega$.

This transformation will act within a space of
 Hamiltonians $H$ of the form
\begin{equation}
\label{eqn:HRG3d}
H(\u{A},\u{\varphi})=\u{\omega}\cdot\u{A}+V(\u{\Omega}\cdot\u{A},\u{\varphi}),
\end{equation}
where $\u{\Omega}\in {\mathbb{R}}^3$ is a vector of Euclidean norm one ($\Vert \u{\Omega}\Vert =1$),
 not parallel to the
frequency vector $\u{\omega}$. We assume that Hamiltonian~(\ref{eqn:HRG3d})
satisfies a non-degeneracy condition ($V^{(2)}_{\u{0}}$ is non-zero).\\
\indent The renormalization is defined by the rescaling and elimination steps.

{\bf (1)} \textit{Rescaling}~: The first part of the transformation is
 composed by a shift of the resonances, a rescaling of time and a
 rescaling of the actions. It acts on a Hamiltonian $H$ as $H'=H\circ
 {\mathcal T}$~:
\begin{equation}
\label{eqn:resc3d}
H'(\u{A},\u{\varphi})=\lambda\mu^{-1} H\left(\frac{1}{\lambda}\tilde{N}\u{A},
N^{-1}\u{\varphi}\right),
\end{equation}
where $\tilde{N}$ denotes the transposed matrix of $N$, and $\lambda$ is the
rescaling coefficient~:
\begin{equation}
\label{eqn:rescaling3d}
\lambda=2\mu^{-1} \Vert N \u{\Omega}\Vert^2 V^{(2)}_{\u{0}}.
\end{equation}
\indent For $H$ given by Eq.\ (\ref{eqn:HRG3d}), this expression becomes
\begin{equation}
H'(\u{A},\u{\varphi})=\u{\omega}\cdot\u{A}+
\sum_{\u{\nu},k} V^{'(k)}_{\u{\nu}}
(\u{\Omega}'\cdot\u{A})^k e^{i\u{\nu}\cdot\u{\varphi}},
\end{equation}
where
\begin{eqnarray}
  \label{eq:renexp3d}
  && \u{\Omega}'=\frac{N\u{\Omega}}{\Vert N\u{\Omega}
  \Vert},\\
  && V^{'(k)}_{\u{\nu}}=r_k
  V_{\tilde{N}\u{\nu}}^{(k)} \; \mbox{ with }
  r_k=2^{1-k}\mu^{k-2}\Vert N\u{\Omega}\Vert^{2-k}\left( V_{\u{0}}^{(2)}\right)^{1-k}.
\end{eqnarray}
\indent We notice that
the frequency vector $\u{\omega}$ of the torus is kept fixed by the
transformation since from Eq.~(\ref{eqn:resc3d}), we have
$\lambda\mu^{-1}\u{\omega}\cdot\frac{1}{\lambda}
\tilde{N}\u{A}=\u{\omega}\cdot\u{A}$.

{\bf (2)} \textit{Elimination}~: The second step is a canonical
transformation ${\mathcal U}$ that eliminates the non-resonant modes of
the perturbation in $H'$.\\
\indent Following Ref.~\cite{chan00d}, we consider the
set $I^-$ of non-resonant modes to be the set of integer vectors
$\u{\nu}\in{\Bbb Z}^3$ such that $\vert \u{\omega}\cdot\u{\nu}\vert
\geq \frac{1}{\sqrt{2}} \Vert \u{\omega}\Vert \Vert \u{\nu}\Vert$.
The canonical transformation ${\mathcal U}$ is such that $H''=H'\circ
{\mathcal U}$ does not have any non-resonant mode, i.e.\ it is
defined by the following equation~:
\begin{equation}
  \label{eq:proj3d}
  {\mathbb{I}}^-(H'\circ {\mathcal U})=0.
\end{equation}
\indent We solve Eq.~(\ref{eq:proj3d}) by a Newton's method following
Ref.~\cite{chan00d}, and we apply the procedure acting on a space of Hamiltonians quadratic in the
actions. \\

These two procedures define the renormalization transformation acting
on a space of Hamiltonians~(\ref{eqn:HRG3d}) with three degrees of freedom
as $H''={\mathcal R}H=H\circ {\mathcal T} \circ{\mathcal U}$. The main
point is that in this renormalization the frequency vector $\u{\omega}$ is kept fixed at each iteration of the
transformation.

\section{Numerical results}
\label{sec:num}

\indent We consider Escande's paradigm model~\cite{esca85} which is a forced 
pendulum model~:
\begin{equation}
  \label{eq:fp}
  H_{\varepsilon}(p,x,t)=\frac{1}{2}p^2-\varepsilon (\cos x+\cos(x-t)).
\end{equation}
\indent We map this Hamiltonian into a time-independent Hamiltonian with two
degrees of freedom of the form~(\ref{eqn:HRG}), or into a
time-independent Hamiltonian with three degrees of freedom of the
form~(\ref{eqn:HRG3d}). We compute the critical coupling at which the
torus is broken by the two renormalization methods $\mathcal{R}$.
We determine $\varepsilon_c$ by iterating the renormalization
transformation on $H_{\varepsilon}$. If the coupling $\varepsilon$ is smaller than a
critical value $\varepsilon_c$, the iterations converge to some integrable Hamiltonian $H_0$
(the coefficients $V_{\u{\nu}}$ for $\u{\nu}\not= \u{0}$ vanish), and if $\varepsilon$
is larger than $\varepsilon_c$, the iterations diverge~:
\begin{eqnarray*}
  && {\mathcal R}^nH_{\varepsilon}\to H_0(\u{A}) \quad \mbox{ for }
  |\varepsilon|< \varepsilon_c,\\
  && {\mathcal R}^nH_{\varepsilon}\to \infty \quad \mbox{ for }
  |\varepsilon|> \varepsilon_c,
\end{eqnarray*}
as $n$ tends to infinity. For numerical convenience, each renormalization $\mathcal{R}$
is defined for quadratic Hamiltonians (all the iterations ${\mathcal{R}}^n
H_{\varepsilon}$ are quadratic in the actions), and at each step,
we truncate the Fourier
series by neglecting all the modes $\u{\nu}=(\nu_i)_{i=1,2 \mbox{ or }
  i=1,2,3}$ such that $\max_i|\nu_i| > L$. Other way of truncating the 
Fourier series can be chosen. The one chosen here is the simplest choice for the 
numerical implementation. \\
\indent Since the renormalization for three degrees of freedom involves more 
Fourier coefficients (proportional to $L^3$, compared to $L^2$ for the 
renormalization for two degrees of freedom), we expect the resulting 
transformation to be much slower than the renormalization for two degrees of 
freedom. However, it turns out from numerical computations that in order to 
reach a given accuracy, both methods require approximately the same amount of 
computational time, i.e.\ the cut-off parameter $L$ for three degrees of freedom 
is chosen smaller than the one for two degrees of freedom. In what follows, the 
thresholds are computed with approximately the same amount of computational 
time.\\   
\indent We apply the above 
procedures for three different frequencies~: the spiral mean, the   Tribonacci 
mean, and the $\tau$-mean frequencies.

\subsection{spiral mean torus}
\label{sec:spiral}

\indent We consider an invariant torus with frequency $\omega=\sigma^{-1}$
where $\sigma$ is the spiral mean~\cite{kim86}, i.e.\ it is the real solution of
$\sigma^3=\sigma+1$. It is approximately $\sigma\approx 1.32472$. The
frequency $\sigma^{-1}$ belongs to the complex cubic field with
negative discriminant $-23$.    The break-up of KAM tori with spiral mean 
frequency of Hamiltonian systems with two degrees of freedom has been previously 
studied using scaling law analysis of Greene's residues of nearby elliptic 
periodic orbits~\cite{mao89b}.  \\

\paragraph*{\textit{Renormalization for  }
$\u{\omega}=(\sigma^{-1},-1)$}  The first entries of the continued fraction 
expansion of $\sigma^{-1}$ are $[1,3,12,1,1,3,2,3,2,4,2,141,80,2,5,\ldots]$.
For a cut-off parameter $L=25$, the renormalization defined in Sec.~\ref{sec:2d} gives
$\varepsilon_c\approx 0.0155$.         \\

\paragraph*{\textit{Renormalization for  }
  $\u{\omega}=(\sigma^{-1},-1,\sigma^{-2})$ }  
The frequency vector $\u{\omega}=(\sigma^{-1},-1,\sigma^{-2})$
is an eigenvector of the matrix
$$
N=\left( \begin{array}{ccc} 0 & 0 &1 \\ -1 & 0 & 0 \\ 0 & -1 & -1
  \end{array} \right).
$$
\indent For a cut-off parameter $L=15$, applying the renormalization defined in
Sec.~\ref{sec:3d} with the above matrix $N$, we obtain $\varepsilon_c\approx 0.016$. If we choose
$\omega_1=\sigma^{-2}+\sigma^{-1}+1$ instead of $\omega_1=\sigma^{-2}$, the
renormalization defined by the matrix
$$
N=\left( \begin{array}{ccc} -1 & -1 & 1 \\ 1 & 0 & -1 \\ 1 & 0 & 0
  \end{array} \right),
$$
gives the same result ($\varepsilon_c\approx 0.016$)
as the renormalization defined for $\omega_1=\sigma^{-2}$.\\
\indent Up to the numerical precision, the renormalizations for two and three 
degrees of freedom give the same result for the critical
threshold $\varepsilon_c$.

\subsection{Tribonacci torus}
\label{sec:tribo}

We consider an invariant torus with frequency $\omega=\varsigma^{-1}$
where $\varsigma$ is called the Tribonacci number and is the real
solution of $\varsigma^3=\varsigma^2+\varsigma+1$. Its value is
approximately $\varsigma\approx  1.83928$. The frequency $\varsigma^{-1}$
belongs to the complex cubic field with discriminant $-31$.       \\

\paragraph*{\textit{Renormalization for }$\u{\omega}=(\varsigma^{-1},-1)$}
The first entries of the continued fraction expansion of $\varsigma^{-1}$
are $[1,1,5,4,2,305,1,8,2,1,4,6,17,5,1,\ldots]$.
For a cut-off parameter $L=25$, we obtain $\varepsilon_c\approx
0.02186$.                                                            \\

\paragraph*{\textit{Renormalization for   }
  $\u{\omega}=(\varsigma^{-1},-1,\varsigma^{-2})$ }
The frequency vector $\u{\omega}=(\varsigma^{-1},-1,\varsigma^{-2})$
is an eigenvector of the matrix
$$
N=\left( \begin{array}{ccc} 0 & 0 &1 \\ -1 & 0 & 0 \\ -1 & -1 & -1
  \end{array} \right).
$$
\indent For a cut-off parameter $L=10$, the renormalization with the above 
matrix $N$ gives  $\varepsilon_c\approx 0.0219$. We notice
that in this case, a very good agreement between both renormalizations is obtained even
for small cut-off parameters $L$.

\subsection{$\tau$-mean torus}
\label{sec:tau}

We consider an invariant torus with frequency $\omega=\tau^{-1}$
where $\tau$ is the real
solution of $\tau^3=-\tau^2+2\tau+1$. Its value is
$\tau=2\cos(2\pi/7) \approx  1.24698$. The frequency $\tau^{-1}$ belongs
to the real cubic field with positive discriminant 49.             \\

\paragraph*{\textit{Renormalization for } $\u{\omega}=(\tau^{-1},-1)$}
The first entries of the continued fraction expansion of $\tau^{-1}$
are $[1,4,20,2,3,1,6,10,5,2,2,1,2,2,1,\ldots]$.
For a cut-off parameter $L=25$, we obtain $\varepsilon_c\approx
0.01247$.                                                                      
\\
 \paragraph*{\textit{Renormalization for   }
  $\u{\omega}=(\tau^{-1},-1,\tau^{-2})$ }
The frequency vector $\u{\omega}=(\tau^{-1},-1,\tau^{-2})$
is an eigenvector of the matrix
$$
N=\left( \begin{array}{ccc} 0 & 0 &1 \\ -1 & 0 & 0 \\ 1 & -1 & -2
  \end{array} \right),
$$
with eigenvalue $\mu=\tau^{-1}$.
For any cut-off parameter, the critical value we obtain by the renormalization
procedure for this matrix $N$ ($\varepsilon_c\approx 0.48$) is very far from the one obtained by the previous
renormalization. The reason is that the matrix $N$ has not the good
properties. In order to define the procedure, the renormalization
transformation has to contract all the integer vectors in $I^+$ (the complement of $I^-$ in
${\mathbb{Z}}^3$) into
vectors in $I^-$~\cite{koch99}. The spectrum of $N$ is composed by
$\tau^{-1},-(1+\tau^{-1})^{-1},-(\tau+1)$. The shift of the resonances
is a map acting on integer vectors as $\u{\nu}\mapsto N^{-1}
\u{\nu}$. Since there are two eigenvalues of modulus larger than one for $N^{-1}$
($\tau$ and $-1-\tau^{-1}$),
there are two unstable directions for the shift of the resonances.

We notice that for the negative discriminant case, this problem cannot
occur. If there is one eigenvalue of modulus smaller than one, then
the two other complex conjugated eigenvalues are of modulus greater
than one since the determinant of $N$ is $\pm 1$.

\paragraph*{\textit{Renormalization for       }
  $\u{\omega}=(\tau^{-1},-1,\tau+1)$ }
The frequency vector $\u{\omega}=(\tau^{-1},-1,\tau+1)$
is an eigenvector of the matrix
$$
N=\left( \begin{array}{ccc} 2 & -1 & -1 \\ 1 & -1 & -1 \\ 0 & -1 & 0
  \end{array} \right),
$$
with eigenvalue $\mu=(1+\tau)^{-1}$.
Here the matrix $N$ has the good property of having only one
eigenvalue of modulus smaller than one (the two other eigenvalues are
$-\tau$ and $1+\tau^{-1}$). The renormalization defined in Sec.~\ref{sec:3d},
for a cut-off parameter $L=7$,
gives $\varepsilon_c\approx 0.013$.
For this frequency vector, an
approximate renormalization has been set up for degenerate
Hamiltonians with three degrees of freedom in Ref.~\cite{chan00b} in
order to investigate self-similar properties of three-dimensional
critical tori. It is defined from the above matrix $N$ and
structurally stable dynamics has been found for this approximate
renormalization. \\

{\em Remark}~: For the forced pendulum model~(\ref{eq:fp}), the
Tribonacci torus is the most robust between the three invariant tori
investigated in this paper, and the spiral mean torus is more
robust than the $\tau$-mean torus. However, this feature
depends on the perturbation in a way that is not understood. The most robust 
invariant tori are conjectured to be noble tori in general (see 
Ref.~\cite{mack92b} for area-preserving maps and Ref.~\cite{chan02a} for the
stochastic ionization of Hydrogen driven by microwaves). Therefore, tori with 
cubic frequency should be less important for the large-scale diffusion of 
trajectories. However, the results presented in this article would be 
useful in order to investigate the break-up of invariant tori in Hamiltonian 
with three degrees of freedom and its comparison with the break-up in two 
degrees of freedom.

\section{Conclusion}
\label{sec:concl}

We have constructed and studied numerically two renormalization
transformations in order to compute thresholds of break-up of invariant
tori with a cubic frequency in Hamiltonian systems with two degrees of
freedom. The main advantage of the procedure acting in a space of
degenerate Hamiltonians with three degrees of freedom, is that it is
independent of the continued fraction algorithm and it is
defined for a fixed frequency vector. Both methods allow an accurate
computation of critical couplings.

\section*{Acknowledgments}
We acknowledge useful discussions with G.\ Benfatto, K.\ Briggs, G.\ Gallavotti,
H.R.\ Jauslin, H.\ Koch, J. Laskar, and
R.S.\ MacKay. Financial support from the Carnot Foundation is also
acknowledged.


\end{document}